\documentclass{aa}  
\usepackage{natbib}
\bibpunct{(}{)}{;}{a}{}{,}
\usepackage[varg]{txfonts}
\usepackage[colorlinks,allcolors=blue]{hyperref}
\usepackage{threeparttable}
\hypersetup{
    colorlinks=true,
    linkcolor=blue,
    filecolor=magenta,      
    urlcolor=blue,
    citecolor=blue
}
\usepackage{ulem}
\usepackage{orcidlink}
\usepackage{hyperref} 
\usepackage{url} 

\begin{document} 
\title{A robust morphological classification method for galaxies using dual-encoding contrastive learning and multi-clustering voting on JWST/NIRCam images}

    \author{Xiaolei Yin\orcidlink{0009-0004-5136-0951}\inst{1,2}       
        \and
        Guanwen Fang\orcidlink{0000-0001-9694-2171}\inst{1,2}       
        \and
        Shiying Lu\orcidlink{0000-0001-5988-2202}\inst{1,2,3}        
        \and
        Zesen Lin\orcidlink{0000-0001-8078-3428}\inst{4} 
        \and
        Yao Dai\orcidlink{0000-0002-4638-0235}\inst{5,6}
        \and
        Chichun Zhou\orcidlink{0000-0002-5133-2668}\inst{7}}
    \institute{
    School of Mathematics and Physics, Anqing Normal University, Anqing 246133, China, \email{wen@mail.ustc.edu.cn}    
    \and Institute of Astronomy and Astrophysics, Anqing Normal University, Anqing 246133, China
    \and Key Laboratory of Modern Astronomy and Astrophysics (Nanjing University), Ministry of Education, Nanjing 210093, China
    \and Department of Physics, The Chinese University of Hong Kong, Shatin, N.T., Hong Kong S.A.R., China
    \and Shanghai Astronomical Observatory, Chinese Academy of Sciences, 80 Nandan Road, Shanghai 200030, China
    \and School of Astronomy and Space Science, University of Chinese Academy of Sciences, No. 19A Yuquan Road, Beijing 100049, China
    \and School of Engineering, Dali University, Dali 671003, China}

\date{Received ---;  accepted ---}

 \abstract
{The two-step galaxy morphology classification framework {\tt USmorph} successfully combines unsupervised machine learning (UML) with supervised machine learning (SML) methods. To enhance the UML step, we employed a dual-encoder architecture (ConvNeXt and ViT) to effectively encode images, contrastive learning to accurately extract features, and principal component analysis to efficiently reduce dimensionality.
Based on this improved framework, a sample of 46,176 galaxies at $0<z<4.2$, selected in the COSMOS-Web field, is classified into five types using the JWST near-infrared images: 33\% spherical (SPH), 25\% early-type disk (ETD), 25\% late-type disk (LTD), 7\% irregular (IRR), and 10\% unclassified (UNC) galaxies. 
We also performed parametric (S{\'e}rsic index, $n$,and effective radius, $r_{\rm e}$) and nonparametric measurements (Gini coefficient, $G$, the second-order moment of light, $M_{\rm 20}$, concentration, $C$, multiplicity, $\Psi$, and three other parameters from the MID statistics) for massive galaxies ($M_*>10^9 M_\odot$) to verify the validity of our galaxy morphological classification system. 
The analysis of morphological parameters is consistent with our classification system: SPH and ETD galaxies with higher $n$, $G$, and $C$ tend to be more bulge-dominated and more compact compared with other types of galaxies.
This demonstrates the reliability of this classification system, which will be useful for a forthcoming large-sky survey from the Chinese Space Station Telescope.}

\titlerunning{A robust morphological classification method for galaxies }
\authorrunning{Yin et al.}
\maketitle

\section{Introduction}\label{sec:1}
The formation and evolution of galaxies have long been a central research topic in the field of astronomy. Studying the physical properties of galaxies is crucial for understanding their formation and evolution \citep{kauffmann+2004, omand+2014, schawinski+2014, kawinwanichakij+2017, Guyizhou+2018, lianou+2019}. The morphological features of galaxies, which are direct observational characteristics, can be obtained through observations from both ground- and space-based telescopes. The morphological characteristics of galaxies across multiple wavelength bands can reveal various physical properties, such as color, dust extinction, star formation activities, and the surrounding environment. With the advent of numerous large-scale sky survey projects, such as the Sloan Digital Sky Survey (SDSS; \citealt{Stoughton+2002}) and the Cosmic Evolution Survey (COSMOS; \citealt{Scoville+2007}), vast amounts of galaxy morphological data have been collected. Classifying galaxies by visual inspection \citep{Hubble+1926, S+1959, Bergh+1976} is inadequate to meet the processing demands of such massive datasets. Therefore, automated morphological classification systems based on machine learning have emerged as an inevitable development \citep{Song+2024,Serrano+2024, Luo+2025}.

Machine learning has become an indispensable tool in galaxy morphology classification and astronomical image analysis, with broad applicability across various tasks \citep{Olaf+2015, He+2016}. In supervised machine learning, convolutional neural networks (CNNs) are powerful tools for image-based analysis \citep {Schmidhuber+2015}. For example,  
\cite {Dieleman+2015} demonstrate the effectiveness of rotation-invariant CNNs on the Galaxy Zoo dataset, setting a precedent for deep learning in galaxy morphology. Subsequent studies, such as \cite{dominguez2018catalog}, demonstrate the scalability of CNNs on large datasets. A series of studies based on the Galaxy Zoo project continue to demonstrate the strong adaptability and effectiveness of CNNs in galaxy morphology classification \citep{Dickinson+2018, Walmsley+2020, Walmsley+2022}. In parallel with these image-based methods, feature-based supervised techniques have also shown promise \citep{McCabe+2021, Reza+2021, rose+2024}. For instance, \cite{rose+2024} use random forest algorithms for feature-based identification of merging galaxies. This approach leverages specific features extracted from the data to classify galaxies, offering an alternative to direct image analysis.

However, effective training of supervised machine learning (SML) models traditionally requires extensive pre-labeled datasets. Despite their strong performance across domains, the reliance of CNNs on labeled data has become a significant bottleneck due to the enormous volumes of data generated by large-scale sky surveys in astronomy. Manually labeling these vast datasets is extremely time-consuming and labor-intensive, restricting the full utilization of machine learning to efficiently analyze and interpret astronomical data.
In contrast, unsupervised machine learning (UML) provides an approach to galaxy morphology clustering that does not depend on preexisting data labels. The UML process involves two main steps: first, features are directly extracted from raw images, and galaxies are then clustered based on these features, grouping those with similar characteristics together. This method allows researchers to discover patterns in astronomical datasets without the need for labor-intensive manual labeling.
Unsupervised methods such as autoencoders and clustering algorithms have also gained attention \citep{Martin+2020, Cheng+2021, Tohill+2024}. \cite{Hocking+2018} applied unsupervised clustering and dimensionality reduction to derive a data-driven taxonomy of galaxies. \cite{bekki2020generative} propose a framework based on a variational autoencoder (VAE) to extract and generate latent morphological features. Additionally, \cite{Cook+2024} used a hierarchical density-based spatial clustering of applications with noise (HDBSCAN) for galaxy classification, demonstrating the potential of density-based methods. Beyond these approaches, other UML paradigms have also been actively explored in the astronomical domain. For example, self-supervised learning has been used in tasks such as feature representation for astronomical images, autonomously learning valuable feature representations from raw astronomical images \citep{Hayat+2021, Parker+2024, Yang+2025}. Meanwhile, domain adaptation methods play a crucial role in addressing distribution mismatches that frequently arise between different astronomical datasets (e.g., \citealt{Ciprijanovic+2023, Belfiore+2025, Ye+2025}). By aligning the feature distributions of the source and target domains, domain adaptation enables models trained on one dataset to generalize more effectively to another, enhancing the robustness and transferability of models across heterogeneous astronomical data sources. These studies highlight how UML serves as an essential approach for handling large quantities of unlabeled astronomical data, mitigating the limitations of traditional supervised pipelines.
Therefore, in the face of vast amounts of unlabeled astronomical data, UML has become a crucial means of addressing the limitations of SML, promoting more efficient analysis and interpretation of astronomical data.

In our previous work, we conducted a series of studies on galaxy morphology classification. First, \cite{Zhou+2022} developed a UML method that includes a robust multiple clustering model using the bagging technique. Their approach incorporates a convolutional autoencoder (CAE; \citealt{Massey+2010}) to reduce noise in the images, thereby improving data quality. Building on this foundation, \cite{Fang+2023} introduced an adaptive polar coordinate transformation (APCT) technique into an SML model to account for residual data after applying the UML model. The APCT step enhances the rotational invariance of images, further boosting the model's robustness. Subsequently, \cite{Dai+2023} and \cite{Song+2024} propose a two-step galaxy morphological classification framework, {\tt USmorph}, combining UML and SML techniques. These advancements demonstrate the successful development of a two-step galaxy morphology classification system. This system exhibits robustness and reliability, making it suitable for future wide-field surveys. In our proposed framework, the UML component is central, as it operates without human-labeled supervision and with minimal human intervention, extracting high-quality features and generating reliable labeled datasets. The supervised stage extends this result by classifying boundary or ambiguous samples, thereby enhancing the expansion and integrity of the dataset. This two-stage process fully leverages the strengths of both approaches: UML for robust, high-fidelity label generation and SML for effective generalization and classification of edge cases.

In this work, to further refine and expand the capabilities of galaxy morphology classification frameworks, we adopted a dual-encoder architecture (i.e., ConvNeXt and ViT) to encode images and incorporated contrastive learning (CL) and principal component analysis (PCA) to accurately extract image features and efficiently reduce the dimensionality, respectively. Specifically, our approach consists of four components: (1) employing a CAE to reduce image noise and applying APCT to the original image to enhance the model's rotational invariance; (2) adopting dual encoding (ConvNeXt and ViT) contrastive learning and PCA to encode images, extract features, and reduce dimensions; (3) using a bagging multi-model to cluster the images based on a voting algorithm; (4) applying the SML technique to train the labeled images and automatically label those discarded during the clustering agglomerative process. 
We subsequently applied this improved system to galaxies within the redshift range 0 $< z <$ 4.2 in the near-infrared (NIR) band of the COSMOS-Web field. In addition, we measured both parametric and nonparametric morphological properties of these galaxies to validate our classification results. The findings indicate that our classification aligns well with the expected relationships among the morphological parameters, demonstrating the reliability of our classification system.

This paper is organized as follows. Section~\ref{sec:2} introduces the COSMOS-Web project and the selection of galaxy samples. Section \ref{sec:3} describes the improved
galaxy morphology classification framework, including data preprocessing techniques, dual-encoding contrastive learning and PCA, unsupervised techniques based on bagging multi-model voting, and the supervised GoogLeNet algorithm. Section \ref{sec:4} presents the clustering results and tests based on morphological parameters of various galaxy types. Section \ref{sec:5} summarizes the main conclusions and provides an outlook on future research directions. We adopt the AB magnitude system \citep{Oke+1983}, assume a \cite{Chabrier+2003} initial mass function, and use a standard flat $\Lambda$CDM cosmology with parameters $H_0 = 70$ km s$^{-1}$ Mpc$^{-1}$, $\Omega_m = 0.3$, and $\Omega_\Lambda = 0.7$.

\section{Data and sample selection} \label{sec:2}
\subsection{The COSMOS-Web survey} \label{sec:2.1}
The COSMOS-Web survey \citep{Casey+2023} is a pioneering large-scale deep-sky observing program conducted by the James Webb Space Telescope (JWST) during its inaugural observing cycle. The primary objective of the program is to elucidate the origin of the Universe and the processes of early galaxy formation through extensive infrared imaging of well-studied regions. These regions have been extensively observed at multiple wavelengths, covering approximately 0.54 deg$^2$. To achieve its scientific goals, the COSMOS-Web project utilizes the Near-Infrared Camera (NIRCam; \citealt{Rieke+2023}) and the Mid-Infrared Instrument (MIRI; \citealt{Wright+2022}) on JWST. In this survey, NIRCam imaging uses four broadband filters (F115W, F150W, F277W, and F444W), enabling high-resolution observations critical for detecting faint and distant galaxies. Imaging with MIRI employs a single mid-infrared filter (F770W), which extends the wavelength coverage and enhances the detection of obscured star-forming regions and active galactic nuclei.

\subsection{The COSMOS2020 catalog} \label{sec:2.2}
The sample used in this study was constructed from the ``Classic'' COSMOS2020 catalog \citep{Weaver+2022}, which offers a comprehensive set of photometric data across 35 bands, ranging from ultraviolet to NIR. Using this enhanced photometric dataset, \cite{Weaver+2022} calculated various physical attributes of galaxies in the field, including photometric redshifts and stellar masses, by fitting spectral energy distributions (SEDs).

To determine redshifts, two distinct codes were employed: {\tt EAZY} \citep{Brammer+2008} and {\tt  LePhare} \citep{Ilbert+2009}. For the present analysis, the redshifts derived from {\tt  LePhare} were adopted because of their demonstrated superior reliability across the considered magnitude range, as shown in Figure 15 of \cite{Weaver+2022}. The redshift estimates are based on a template library of 33 galaxies described by \cite{Bruzual+2003} and \cite{Ilbert+2009}. Additionally, a variety of dust extinction and attenuation curves were used, including the starburst attenuation curves of \cite{Calzetti+2000}, the SMC extinction curves of \cite{Prevot+1984}, and two variants of the Calzetti law incorporating the 2175 \AA\ UV bump. The photometric redshift was then defined as the median value derived from the redshift likelihood distribution.

\subsection{Sample selection} \label{sec:2.3} 
In the COSMOS-Web project, we selected our sample based on the principle of stellar mass completeness. Additionally, we selected images in the rest-frame NIR band using four infrared filters (F115W, F150W, F277W, and F444W) 
The selection of our parent sample from the COSMOS2020 catalog adheres to the following criteria: (1) $\rm lp_{type}=0$, which ensures that the selected target is a galaxy; (2) $\rm FLAG_{ COMBINE}=0$, which implies that the flux measurements are unaffected by the presence of bright stars and that the objects are in the center of the image, ensuring the reliability of the photometric redshift and mass estimates; (3) $0<z<4.2$, ensuring morphology estimation in the rest-frame NIR band.

\begin{figure}	
\includegraphics[width=1.0\columnwidth]{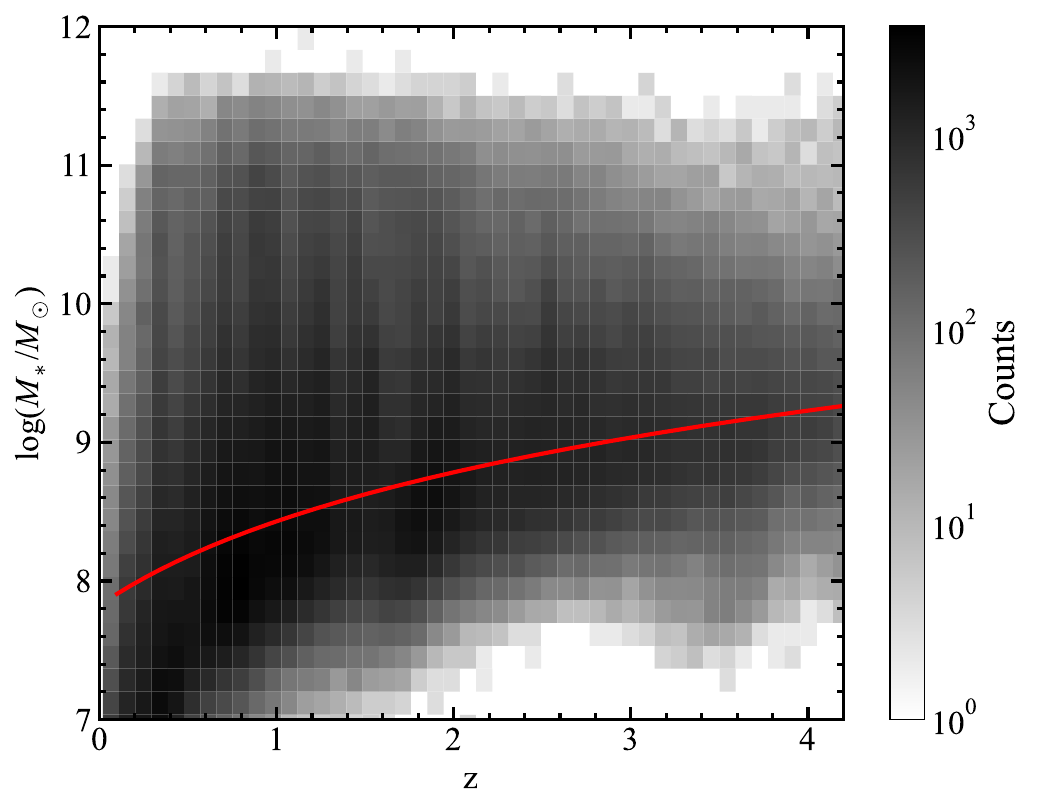}	
\caption{Relationship between stellar mass and redshift. The solid red line shows the stellar-mass completeness, described by $ M_{\rm comp}(z)/M_{\odot} = -1.51 \times 10^6 (1 + z) + 6.81 \times 10^7 (1 + z)^2$. \label{fig1}}
\end{figure}

\begin{table}
\centering
\caption{Galaxy sample in different bands and redshift intervals}\label{tab:1}
\begin{tabular}{c|c|c|c}
\hline\hline
Redshift & Band & ${\lambda}_{\text{rest}}$ (Å) & Total \\
\hline
$0 < z \leq 0.2$ & F115W & 9,997 & 345 \\
$0.2 < z \leq0.6$ & F150W & 10,633 & 5,297 \\
$0.6 < z \leq2.3$ & F277W & 12,410 & 30,384 \\
$2.3 < z < 4.2$ & F444W &11,306 & 10,150 \\
\hline
\end{tabular}
\begin{tablenotes}
\item ${\lambda}_{\text{rest}}$ gives the median wavelength of galaxies in the rest frame across different redshift ranges. 
\end{tablenotes}
\end{table}

In this study, we selected galaxies based on mass completeness to ensure that those analyzed within a specific redshift range are complete and representative. The methodology for mass completeness was adopted from \cite{Weaver+2022}. Specifically, \cite{Pozzetti+2010} defined a minimum stellar mass ($M_{\rm min}$) at each redshift, above which the selected sample is complete. To determine the minimum stellar mass, we estimated the limiting masses of galaxies at different redshifts using the completeness limit function given by $ M_{\rm comp}(z)/M_{\odot} = -1.51 \times 10^6 (1 + z) + 6.81 \times 10^7 (1 + z)^2$. Fig.~\ref{fig1} shows the stellar mass distribution of galaxies in the COSMOS2020 catalog at $0<z<4.2$. The solid red line indicates the completeness limit function, and we selected the galaxies above this line as the parent sample.

Finally, we selected 46,176 galaxies in the COSMOS-Web field. The corresponding redshift intervals for the four filters are summarized in Table \ref{tab:1}, ensuring that the rest-frame NIR morphologies of galaxies are observed at approximately 1.1$\mu$m.

\section{Morphological classification method} \label{sec:3}
\begin{figure*}[ht!]    
\includegraphics[width=2\columnwidth]{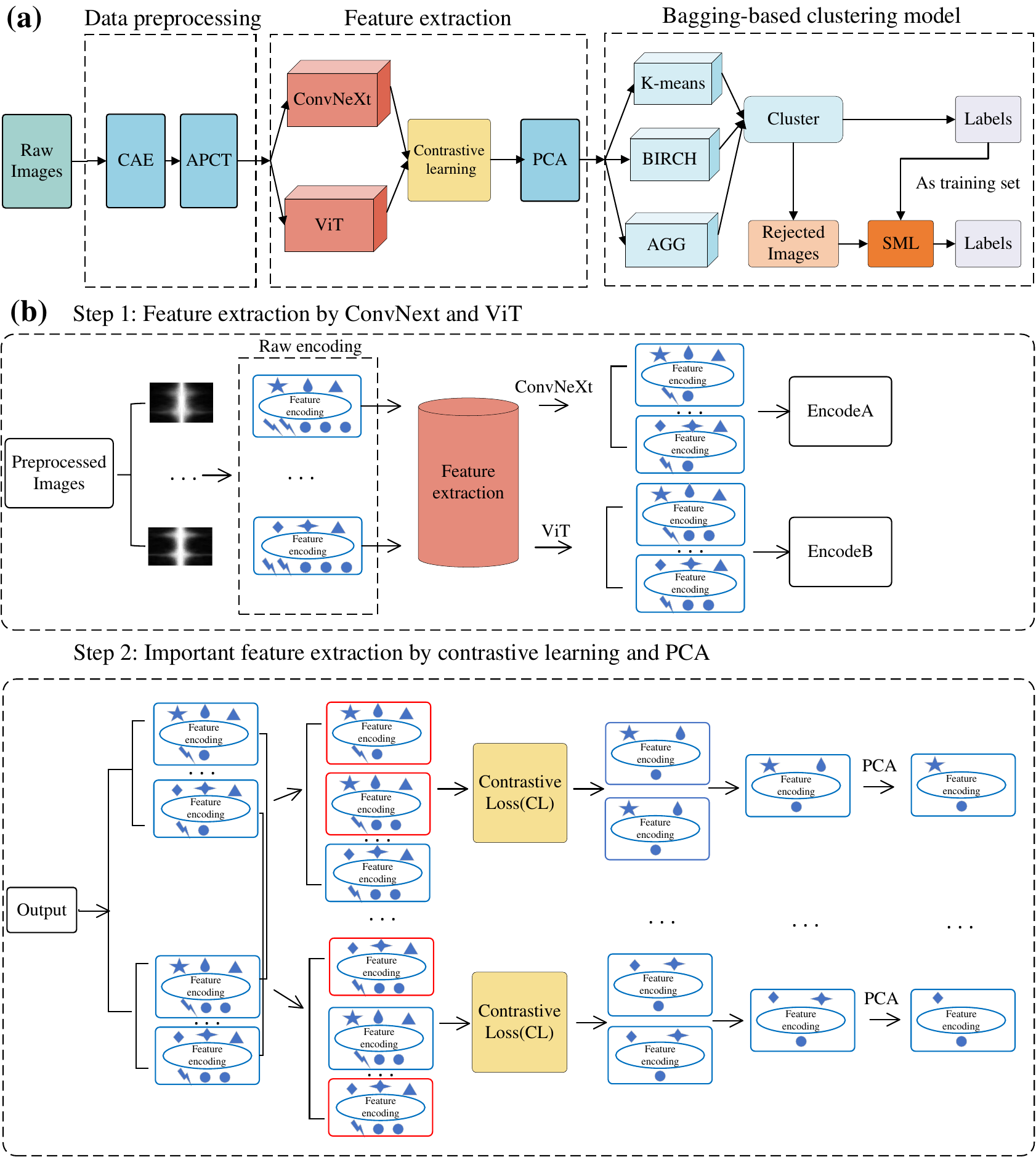}    
\caption{Schematic of the improved galaxy morphology classification system. Panel (a) shows the flow of the galaxy morphology classification system. Panel (b) illustrates feature extraction using ConvNeXt and ViT models to encode the data and extract key features. Encoding results from the same data instances are used to construct positive sample pairs, while the remaining encoding from the ViT model form negative sample sets. The contrastive loss (CL) function is applied to minimize the loss and extract important features. Finally, dimensionality reduction is performed using PCA on the encoded results after contrastive learning.
}    
\label{fig:2}    
\end{figure*}

In this section, we introduce an improved framework that combines dual-encoding contrastive learning and PCA to enhance the performance of our model (as shown in Fig.~\ref{fig:2}). This framework comprises four components: data preprocessing, feature extraction, the UML method, and the SML method.

\subsection{Data preprocessing}
\begin{figure*}[ht!]
\centering	
\includegraphics[width=2\columnwidth]{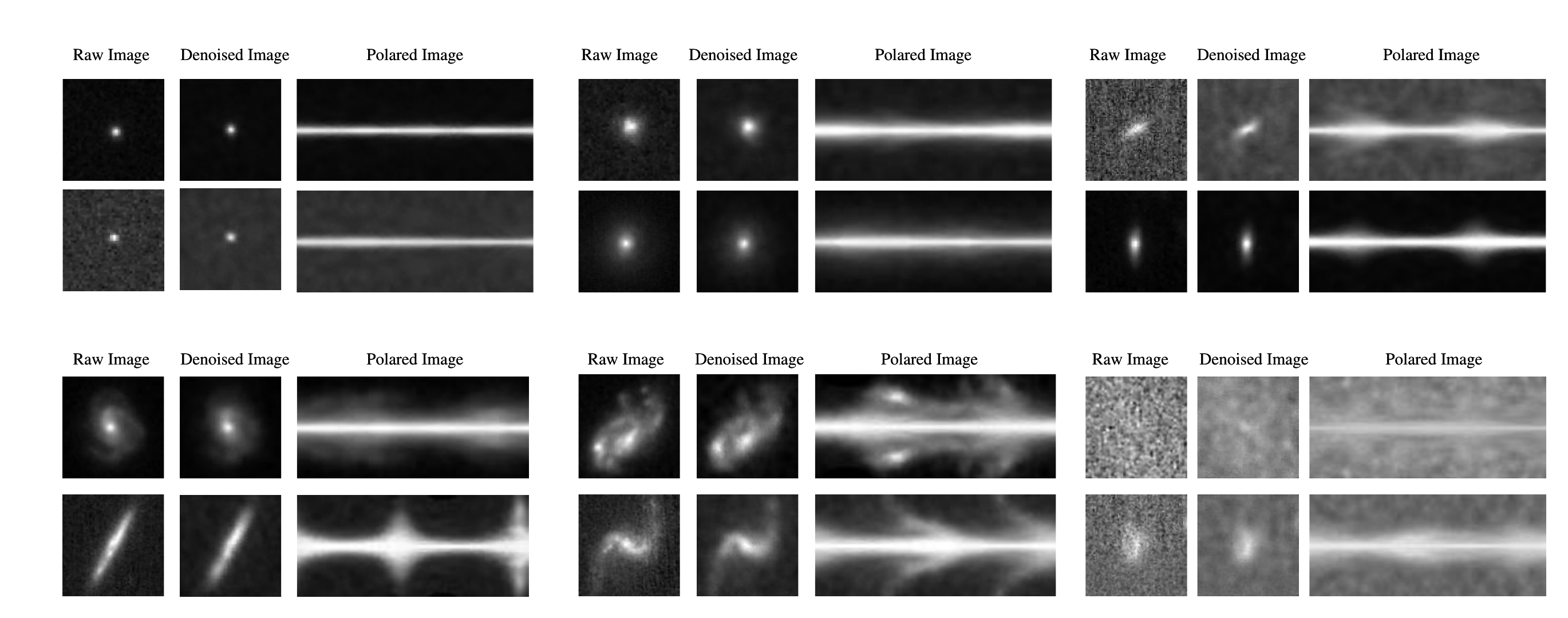}	
\caption{Six sets of images demonstrating image preprocessing steps. In each set, the left, center, and right panels show the original images in the NIR band, post-CAE-based denoised images, and images after polar coordinate expansion, respectively.} \label{fig:fig3}
\end{figure*}

To enhance the stability and accuracy of the classification model, we performed data pre-processing on the galaxy images to reduce image noise and improve rotational invariance. Figure~\ref{fig:fig3} illustrates the differences between raw and preprocessed galaxy images, highlighting the effectiveness of noise reduction and rotational invariance enhancement. To address the adverse effects of high noise levels on classification accuracy, we implemented a CAE-based denoising framework \citep{Massey+2010}. The architecture leverages hierarchical convolutional and pooling operations to autonomously extract latent spatial features from raw galaxy images. By reconstructing denoised representations through learned feature mappings, the CAE suppresses stochastic noise artifacts while preserving critical morphological patterns essential for classification.  As shown in Fig.~\ref{fig:fig3}, each group of images contains three distinct components: the raw image (left panel), the denoised reconstruction generated by the CAE (central panel), and the polar-transformed output (right panel). The visual comparison between raw and denoised images shows that the CAE architecture successfully suppresses stochastic noise while preserving essential morphological features of galaxies. The implementation in this work builds on the framework developed by \cite{Zhou+2022}, which forms the methodological foundation of our research series and demonstrates the effectiveness of CAEs for image denoising in this context.

Some studies have shown that spatial rotation operations can induce misclassification of galaxy morphological types, reducing classification accuracy \citep{Dieleman+2015,Yao+2019}. To address this limitation, we implemented the APCT technique proposed by \cite{Fang+2023}. Their research establishes that APCT-enhanced CNNs effectively improve classification precision while substantially enhancing model robustness against orientation variations, particularly in scenarios requiring rotational invariance. This workflow begins by defining the initial polar axis based on extremal luminance points (maximum and minimum pixel values) within the image. The polar axis is then rotated counterclockwise in angular increments of 0.05 radians. At each rotational position, pixel intensities along the current polar axis direction are integrated through polar coordinate projection. A mirroring operation is then applied to amplify centrally symmetric features. This pipeline enhances the representational capacity of galaxy morphological structures while improving model rotational invariance via coordinate system transformation. As demonstrated in the right panels of Fig.~\ref{fig:fig3}, the APCT-processed images exhibit accentuated critical morphological features such as spiral arm structures.


\subsection{Feature extraction, contrastive learning, and PCA} \label{sec:3.2}
Pre-trained visual classification models are typically trained based on a wide range of generalized image datasets, resulting in efficient initial parameters and strong feature representations. This approach allows the models to extract key features even in the absence of input data labels. The models can then be adapted to specific task requirements through targeted fine-tuning. State-of-the-art models such as ConvNeXt \citep{Liu+2022} and ViT \citep{Dosovitskiy+2020} have proven highly effective as feature encoders, demonstrating excellent performance in diverse image classification applications \citep{liu+2023, Fernandes+2023,Deng+2024,He+2025}.
Contrastive learning is a self-supervised method that builds feature representations by learning similarities and differences between data samples. Specifically, this approach compares positive sample pairs (e.g., different transformations of the same image) and negative sample pairs (e.g., different images), mapping positive pairs close together in feature space while pushing negative pairs apart. Effective contrastive learning requires high-quality positive and negative sample pairs. Improper selection of negative samples can cause the model to learn incorrect feature representations, affecting the final performance.

Traditional contrastive learning typically applies a series of transformations to the original data (such as cropping, flipping, and color jittering) to generate positive sample pairs. Designing an effective data augmentation strategy that not only retains semantic information but also enhances model robustness is challenging \citep{Chen+2020b,Chen+2020a,Grill+2020}. Our study employs a novel sample modeling framework based on encoded outputs from pretrained large-scale supervised models, including ConvNeXt and ViT. This framework leverages the powerful feature representation capabilities of pretrained models to effectively reduce reliance on prior knowledge of data augmentation and enhances the robustness of feature extraction, thereby improving adaptability to a wider range of datasets. Specifically, we used ConvNeXt and ViT to encode each sample in parallel (see Fig.~\ref{fig:2}b Step 1). ConvNeXt outputs features with a dimensionality of 2048, while ViT produces features of 1000 dimensions. To align the feature spaces, an independent linear projection layer maps the output of ViT to a 2048-dimensional feature space. This unified feature representation preserves the semantic information from the original models and provides a measurable feature space for comparative analysis of positive and negative sample pairs, as shown in the following equations:
\begin{equation}    
EncodeA = ConvNeXt(x_1, x_2, x_3, ... , x_n);
\end{equation}
\begin{equation}    
EncodeB = Linear(ViT(x_1, x_2, x_3, ... , x_n)),
\end{equation}
where $X$\(_{(x_1, x_2, x_3, ... , x_n)}\) denotes the set of pixel matrices of the input images, \(ConvNeXt( )\) and \(ViT( )\) denote the encoder type, and \(Linear( )\) denotes the linear layer.

After encoding, we constructed positive and negative sample pairs. To enhance contrastive learning, we used the encoding results of the less effective feature encoder as the negative sample set. ConvNeXt exhibits superior feature extraction capabilities compared to ViT, as indicated by \cite{liu+2023} and \cite{huang+2025}.
Therefore, for each galaxy, the encoding results of ConvNeXt (EncodeA) and ViT (EncodeB) were used as positive sample pairs, with the output from ConvNeXt serving as the anchor sample and the output from ViT serving as the positive sample. The encoding results of all other galaxies from ViT (EncodeB) served as the negative sample set.
Each positive sample pair from EncodeA and EncodeB was matched against all negative sample encodings in EncodeB for contrastive learning. A nonlinear layer mapped the sample encodings to key features under the guidance of the contrastive learning loss function.  After the nonlinear layer completed the encoding of positive and negative samples, performance was evaluated using contrastive loss (CL). The smaller the CL, the closer the output of the nonlinear layer is to the key features of the target data. This brings similar samples closer together in feature space while pushing dissimilar samples apart. The contrastive loss (CL) is defined as
\begin{equation}    
CL = \frac{1}{2n} \sum_{i=1}^{n} \text{loss}_i,
\end{equation}
where $n$ is the number of input data, $ \text{loss}_i $ is the loss function between each pair of positive and negative samples, and $ i $ denotes the data selected as an anchor sample and positive sample from the $n$ data processed by Eqs. (1) and (2). The definition of $ \text{loss}_i $ is as follows:
\begin{equation}    
\text{loss}_i = - \log \frac{\exp(\text{sim}(x_i, y_i) / \tau)}{\exp(\text{sim}(x_i, y_i) / \tau) + \sum_{j \neq i} \exp(\text{sim}(x_i, y_j) / \tau)},
\end{equation}
where $ x_i $ is the anchor sample from Equation (1), $ y_i $ is the positive sample from Equation (2), and $ j \neq i $, $ y_j $ represents the negative sample of the selected data after processing. The function $ \text{sim}() $ denotes the cosine similarity between two different samples, defined as 
\begin{equation} 
\text{sim}(x_i, y_j) = \frac{z_i \cdot z_j}{\|z_i\| \cdot \|z_j\|},
\end{equation}
where $ z_i \cdot z_j $ is the dot product between two different samples and $ \|z_i\| \cdot \|z_j\| $ represents the product of their magnitudes.

Principal component analysis (PCA; \citep{MACKIEWICZ+1993}) is a widely used linear transformation method for dimensionality reduction and feature extraction. Its core idea is to project high-dimensional data into a lower-dimensional space through an orthogonal transformation while retaining the primary variance information in the data. Specifically, PCA computes the covariance matrix of the data and determines its eigenvectors. It then selects the top $k$ eigenvectors associated with the largest eigenvalues as principal components, which are used to project the original data into a new $k$-dimensional space. Typically, the $k$ values should be chosen such that the selected eigenvalues account for the vast majority of the data variance, thus ensuring that most of the original data information is retained. This approach is effective in reducing dimensionality while retaining the essential features of the dataset. In this study, we applied PCA to reduce the dimensionality of the original 1,024-dimensional feature vectors, aiming to eliminate redundant information, reduce computational complexity, and preserve the primary structural characteristics of the data. Through this process, we extracted key features more efficiently, providing a more concise and representative data representation for subsequent analysis and modeling.

As shown in panel b of Fig.~\ref{fig:2}, dual-encoding contrastive learning based on outputs from ConvNeXt and ViT models extracts meaningful features from unlabeled data by constructing positive and negative sample pairs (Step 1). Meanwhile, PCA, as a dimensionality reduction technique, reduces redundancy and noise by maximizing the variance direction of the data. By integrating these two approaches, high-quality feature representations are first acquired using contrastive learning, followed by PCA for dimensionality reduction to simplify the model and improve computational efficiency (Step 2). This method not only retains critical features, but also facilitates data visualization. Overall, the combination of contrastive learning and PCA optimizes feature representation, provides refined and effective input for downstream classification tasks, and highlights the complementary advantages of both techniques in feature extraction.

\begin{figure*}[htpb]
\includegraphics[width=2\columnwidth]{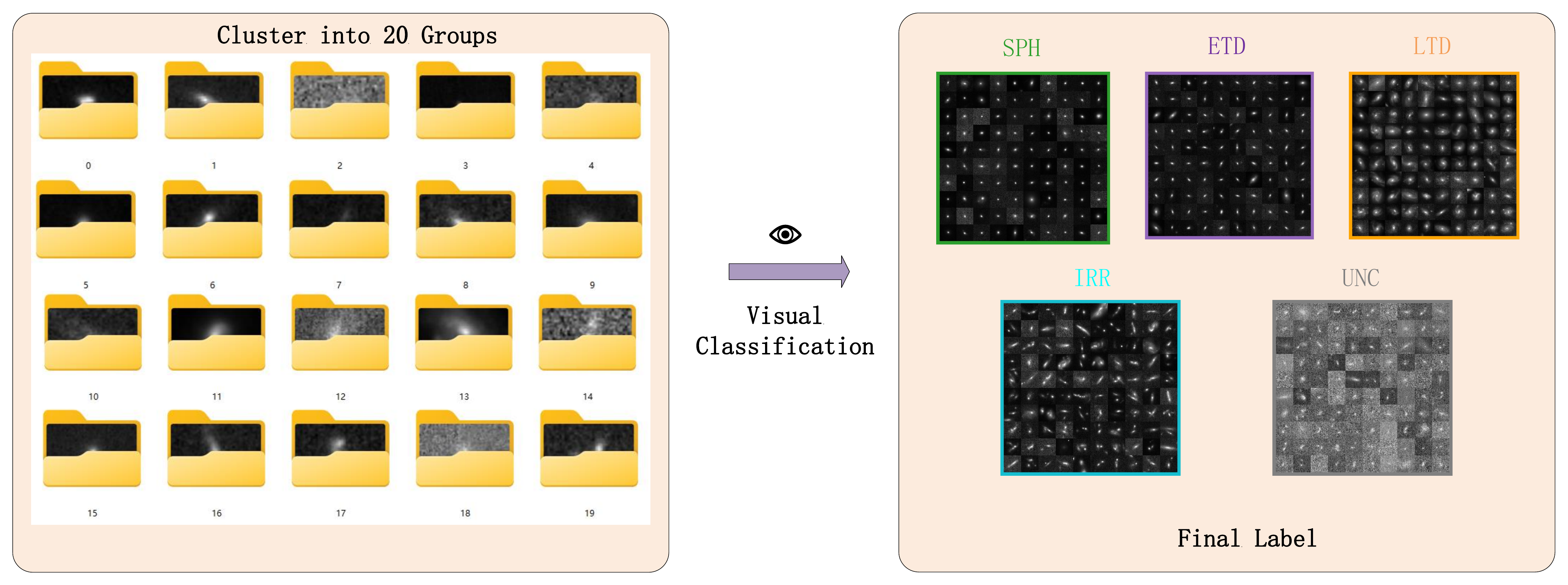}   
\caption{Random selection of 100 images for each of five galaxy types, identified from 20 groups through visual inspection. The clustering results show small morphological differences within groups and large differences between groups, enabling visual classification of each component into SPH, ETD, LTD, IRR, and UNC types.}    
\label{fig:4}
\end{figure*}
\begin{figure*}	[ht!]
\includegraphics[width=2\columnwidth]{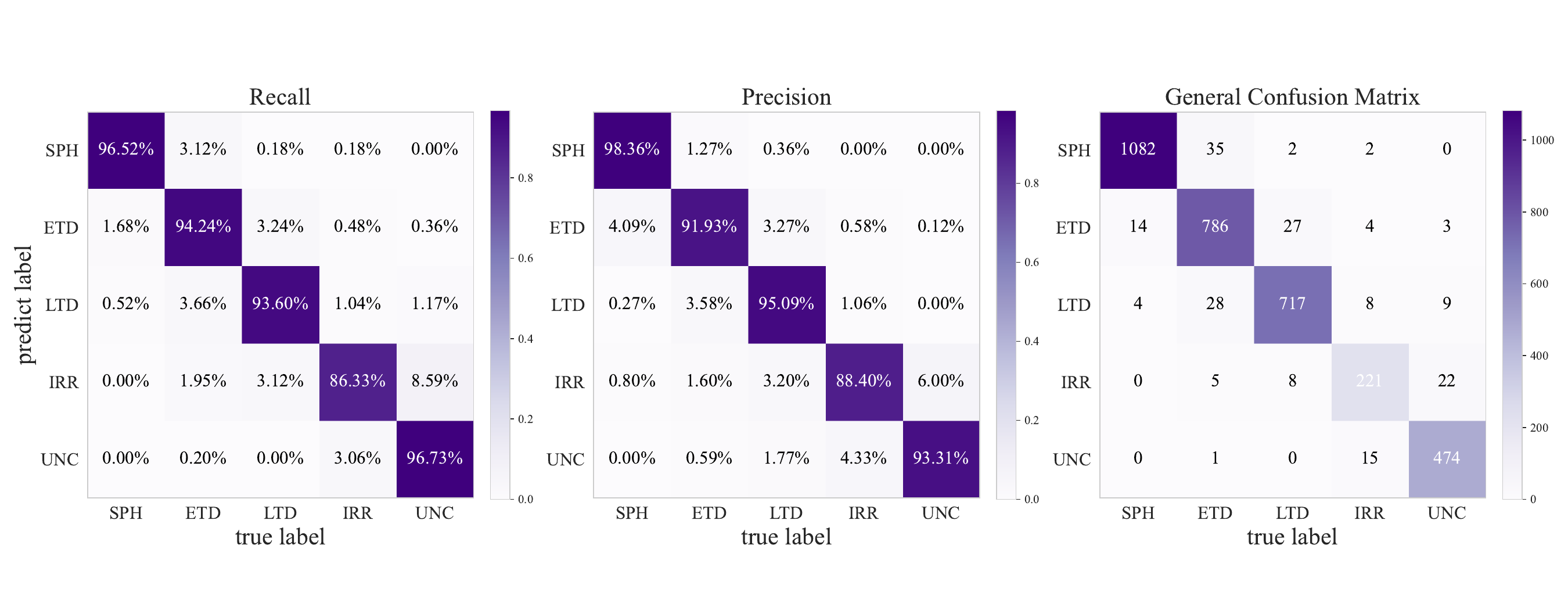}    
\caption{Galaxy morphology classification performance: recall, precision, and general confusion matrix analysis. Left and middle panels represent the recall and precision of the GoogLeNet model, with both exceeding 94.6\%, which indicates that the GoogLeNet model can effectively distinguish between different types of galaxies. The right panel represents the general confusion matrix.}    
\label{fig:5}
\end{figure*}

\subsection{UML clustering process} \label{sec:3.3}
After data preprocessing, feature extraction, and dimensionality reduction, we clustered the galaxy sample using the UML method proposed by \cite{Zhou+2022}, which employs a multi-model voting classification method based on bagging. We used three algorithms: balanced iterative reducing and clustering using hierarchies\citep{zhang+1996}, k-means clustering\citep{Hartigan+1979}, and agglomerative clustering \citep{Murtagh+1983,Murtagh+2014}. Each model performed clustering on the samples, partitioning them into 20 distinct groups. In our analysis framework, the cluster labels generated by the k-means algorithm served as reference labels. For each cluster produced by the alternative models, we assigned labels according to the most frequently occurring k-means label within that cluster. This label alignment process was followed by the implementation of a majority voting mechanism to determine the final classification. The final sample set was made up of the intersection of samples where the three algorithms yielded identical voting results. In particular, only those data points that achieved unanimous consensus across all three models were retained in the final collection, while any samples that fell outside this consensus set were systematically discarded. A detailed schematic representation of this process is provided in Figure 4 of \cite{Zhou+2022}.

After excluding 28,850 contentious targets in the parent sample of 46,176 galaxies, a total of 17,326 galaxies were successfully classified into 20 distinct clusters. Each cluster consisted of galaxies that exhibited similar morphological characteristics, with more significant morphological distinctions observed between different clusters. To further our analysis, we conducted a visual classification of 100 randomly selected galaxies from each cluster. Based on the comprehensive morphological features of the samples within each group, these clusters were classified into one of five galaxy types: spherical (SPH), early-type disk (ETD), late-type disk (LTD), irregular disk (IRR), or unclassified (UNC). Finally, all 17,326 galaxies were systematically assigned to one of the five categories, as illustrated in Fig.~\ref{fig:4}.

\subsection{SML classification process} \label{sec:3.4}
After removing 28,850 inconsistent votes, we obtained 17,326 galaxies with reliable morphological labels through the aforementioned UML clustering and visualization process. These galaxies were used as a training set for the SML classification of the remaining 28,850 samples. \cite{Fang+2023} demonstrate that GoogLeNet outperforms several CNN algorithms in galaxy morphology classification, so we adopted GoogLeNet \citep{Szegedy+2015} as the supervised classification model in this article.

To avoid overfitting, we used the same dataset setup as described in \cite{Fang+2023} to divide the 17,326 labeled samples from the UML clustering process into training and validation sets in a fixed ratio of 9:1. The performance metrics of GoogLeNet on the validation set are presented in Fig.~\ref{fig:5}. The model achieves good classification performance using labels obtained through the UML process, with an overall accuracy rate of approximately 94.6\%, as evidenced by both accuracy and recall metrics. The classification results for 28,850 galaxies using the SML method are listed in Table~\ref{tab2}.

\begin{table}
\centering
\caption{Final classification results by morphological type.\label{tab2}}
\resizebox{0.45\textwidth}{!}{
\begin{tabular}{c|ccccc|c}
\hline\hline
Model & SPH & ETD & LTD  & IRR & UNC& TOTAL\\
\hline
UML & 5,497 &4,275 & 3,769 & 1,246& 2,539 & 17,326\\
SML & 9,694 &7,339& 7,670 & 1,875& 2,272 & 28,850\\
\hline
TOTAL& 15,191 &11,614 & 11,439 & 3,121& 4,811 & 46,176\\
\hline
\end{tabular}
}
\end{table}

\begin{figure*}[ht!]
\centering 
\includegraphics[width=0.9\linewidth]{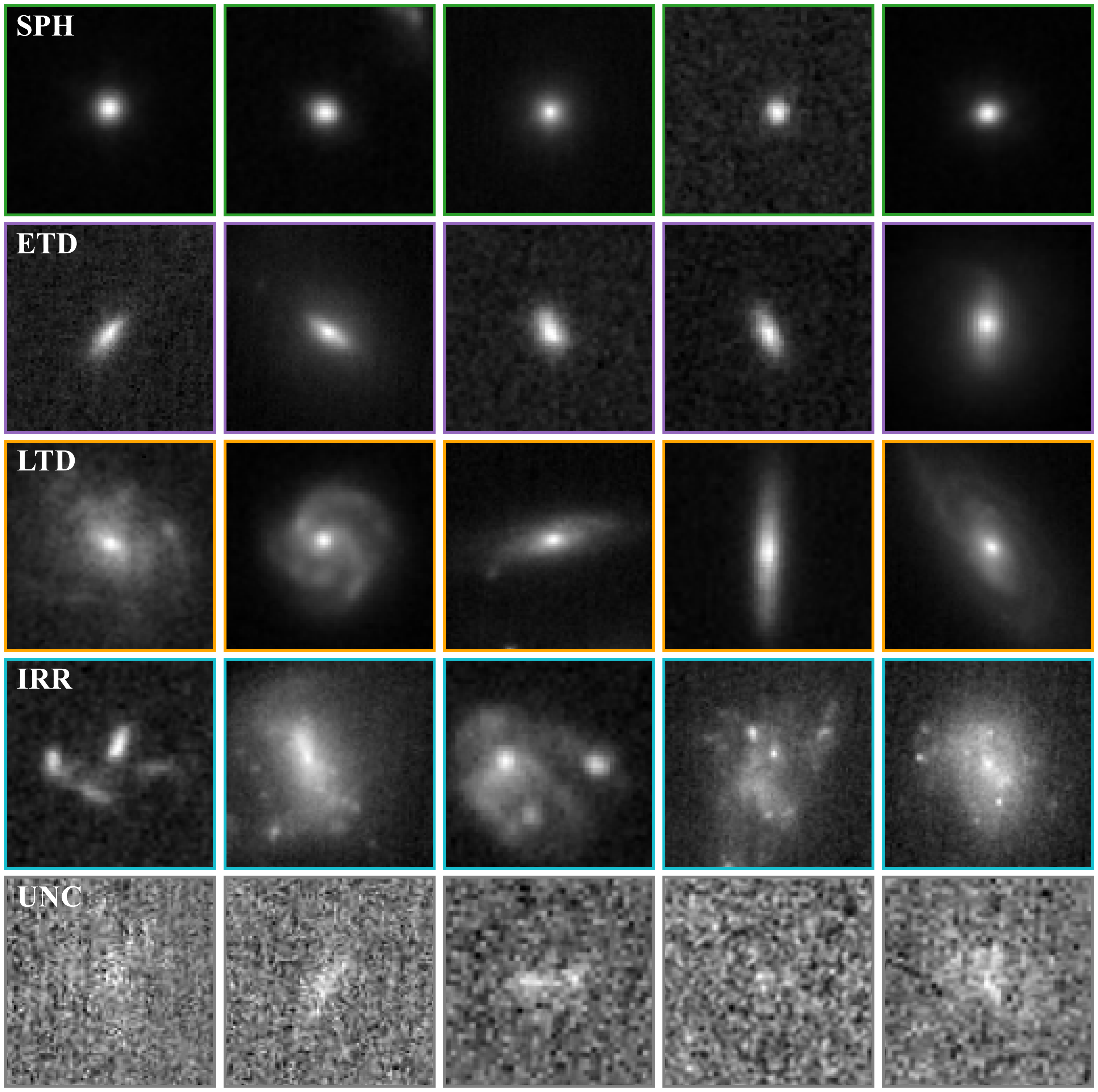}    
\caption{Example NIR-band images of five types of galaxies. The top-to-bottom panels show images from spherical galaxies (SPHs), early-type disk galaxies (ETDs), late-type disk galaxies (LTDs), irregular galaxies (IRRs), and unclassified galaxies (UNCs), respectively. 
}
  
\label{fig:6}
\end{figure*}

\begin{figure*}[ht!]  
\centering    
\includegraphics[width=2\columnwidth]{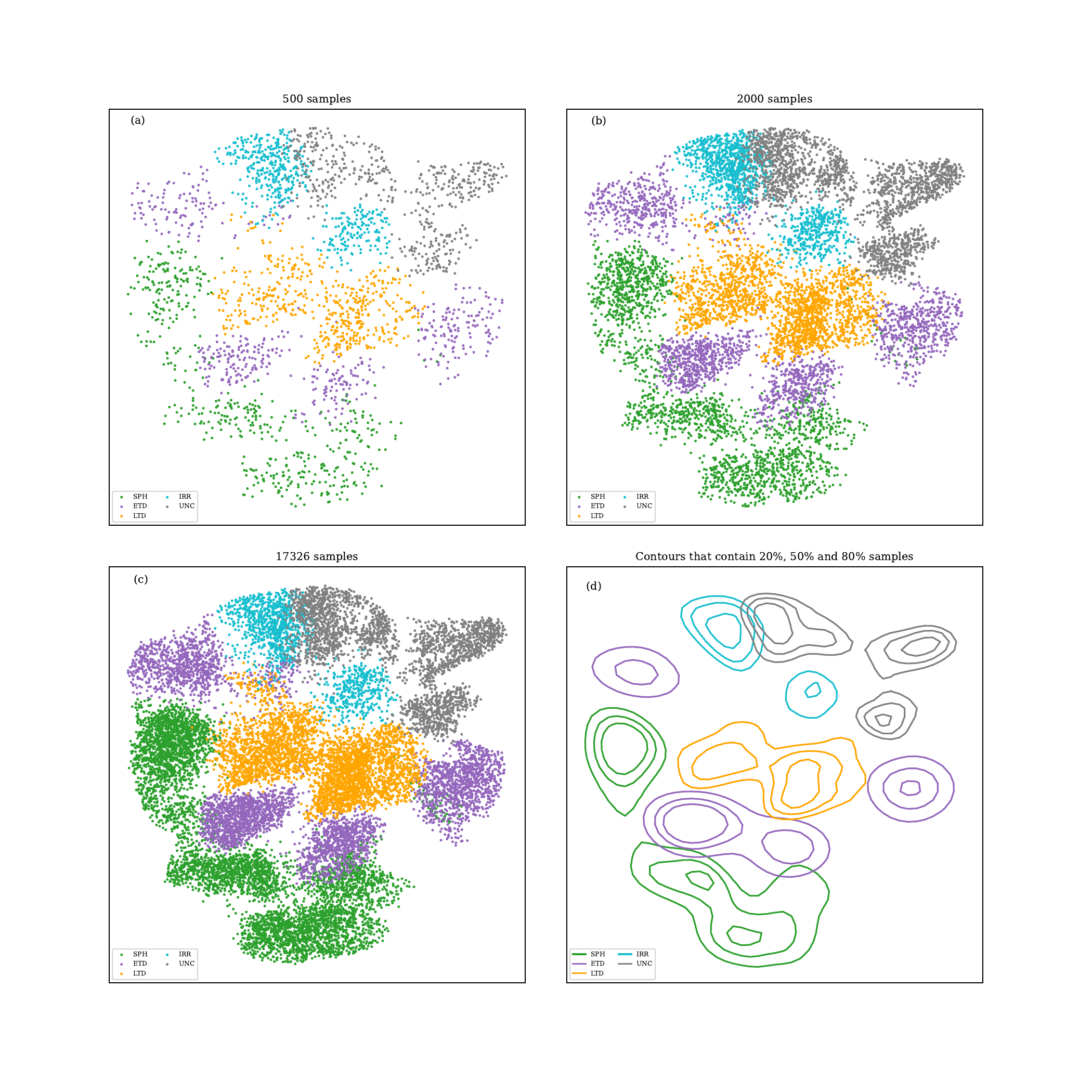} 
\caption{Visualization of the final classification results using UMAP. The UMAP dimensionality reduction technique projects the five classified galaxy types into a 
two-dimensional space. Panel (a) shows the dimensionality reduction results for a random subset of 500 samples, panel (b) for 2000 samples, and panel (c) for the entire dataset. The contours in panel (d) enclose 20\%, 50\%, and 80\% of the corresponding samples.
}    
\label{fig:7}
\end{figure*} 

\begin{figure*}[ht!]    
\includegraphics[width=2\columnwidth]{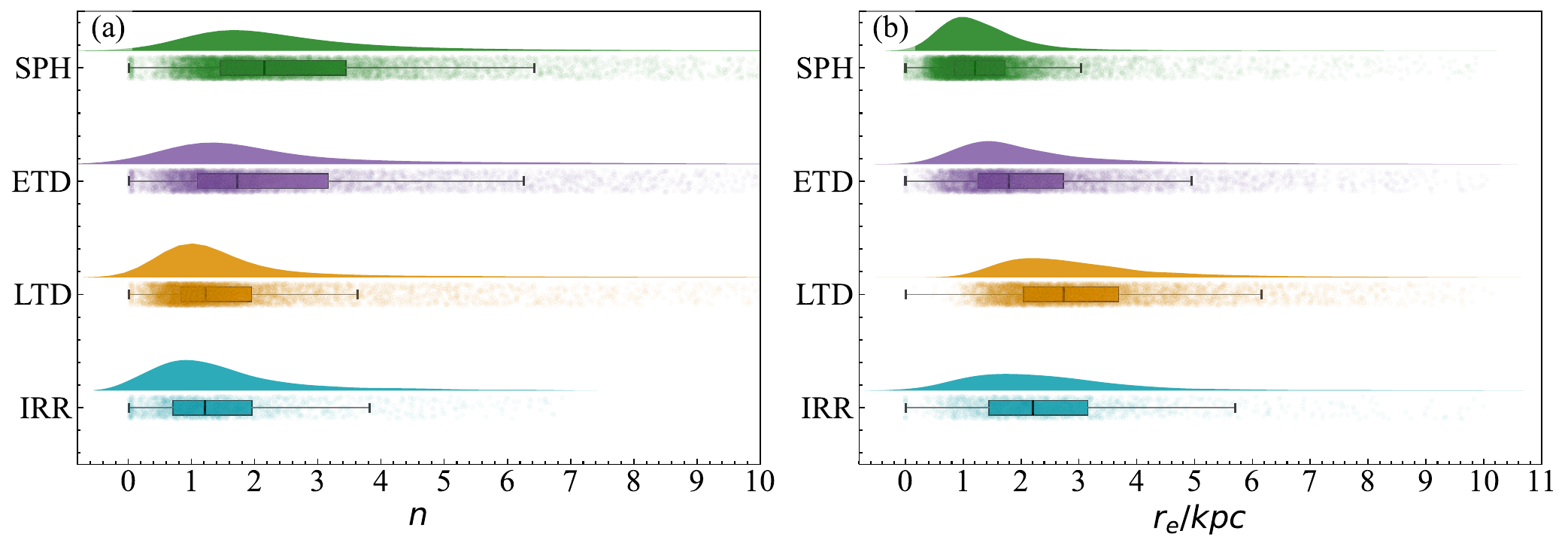}    
\caption{Raincloud plots for different types of massive galaxies: Sérsic index (left) and effective radius (right). ``Cloud'' sections show kernel density distributions, while ``rain'' parts show individual value distributions. In the box plots, bounds denote the first and third quartiles, the central line marks the median, and the whiskers indicate the minimum and maximum values.}    
\label{fig:8}
\end{figure*}

\section{Results and discussion} \label{sec:4}
\subsection{Overall morphological classification results} \label{sec:4.1}
In this study, we successfully classified 46,176 galaxies from the COSMOS-Web field in the NIR band using the updated galaxy morphology classification system, including 15,191 SPHs, 11,614 ETDs, 11,439 LTDs, 3,121 IRRs, and 4,811 UNCs. Detailed classification results are summarized in Table~\ref{tab2}. Figure~\ref{fig:6} shows samples of the five different galaxy types, and the differences between them can be discerned by visual inspection.
Specifically, SPHs exhibit a uniform spatial distribution and a compact morphology.
The ETDs are characterized by a prominent central nucleus and a less compact structure, with both bulges and disks, but lacking well-defined rotating arms. The LTDs have extended structures dominated by a disk, which is usually accompanied by distinct rotating arms and a more dispersed luminosity profile. The IRRs show a variety of morphological features, including asymmetric structures, disordered morphology, and potential merger relics. The UNCs typically have low signal-to-noise ratios (S/Ns) and remain morphologically ambiguous due to their fainter photometric profiles and lower structural resolution, which prevents a reliable classification within our established framework.

In addition, we employed the uniform mobility approximation and projection (UMAP; \citealt{McInnes+2018}) technique to better visualize and evaluate the clustering results. 
As UMAP preserves both the local and global structure of high-dimensional data, it is essential for accurately present our galaxy morphology classification results. Specifically, UMAP can efficiently process large-scale datasets (n = 46,176) while preserving topological relationships, allowing us to visualize intrinsic clustering patterns and interclass relationships in multidimensional feature spaces \citep{McInnes+2018}. Fig. \ref{fig:7}, presents UMAP projections, showing the distribution of our clustering results through three different sample scales: (a) a randomly selected subset of 500 galaxies, (b) an extended sample of 2,000 galaxies, and (c) the full set of 17,326 classified galaxies. The two-dimensional (2D) embedding space reveals the intrinsic structure of the 20 clusters, which are presented in five morphological categories to enhance readability.
Panel (d) of Fig.~\ref{fig:7} shows that more than 80\% of the samples have no overlap in the UMAP-transformed 2D projection map, with clearly discernible cluster boundaries, demonstrating the effectiveness of our unsupervised clustering. However, the UMAP provides only a qualitative overview of the clustering results, and we further validate our classification results quantitatively in the following sections.

\subsection{Test of morphological parameters} \label{sec:4.2}

The structural parameters and nonparametric structural parameters of galaxies are closely related to galaxy morphology, and different galaxy categories typically occupy distinct positions in the parameter space \citep{Lotz+2008,Yao+2023}. Morphological parameters serve as critical diagnostic metrics for evaluating the robustness and reliability of our classification scheme \citep{2021MNRAS.503.4446C, Dai+2023, Song+2024, Tohill+2024}. In this section, we conduct a systematic validation of our classification framework through a comprehensive quantitative analysis of multiple morphological parameters, focusing on massive galaxies characterized by stellar masses $M_{*}>10^{9}~M_\odot$. To ensure the accuracy and reliability of our measurements of physical parameters, galaxies of UNC type are excluded from our analysis because their inherently low S/Ns prevent a reliable determination of morphological parameters.

\begin{figure*}[ht!]	
\includegraphics[width=2\columnwidth]{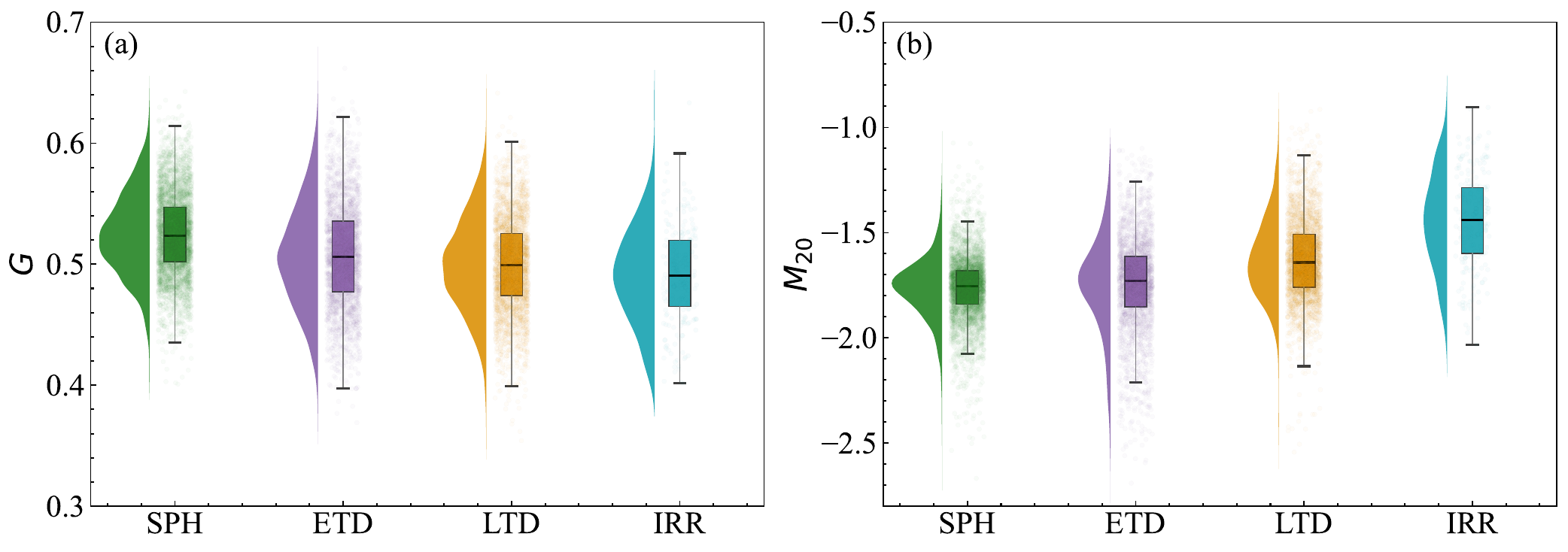}    
\caption{Raincloud plots of G (left) and $M_{20}$ (right) for different types of massive galaxies. Plots are analogous to those in Fig.~\ref{fig:8}. 
From SPH to IRR types, G decreases while $M_{20}$ gradually increases.}    
\label{fig:9}
\end{figure*}
\begin{figure*}[ht!]	
\includegraphics[width=2\columnwidth]{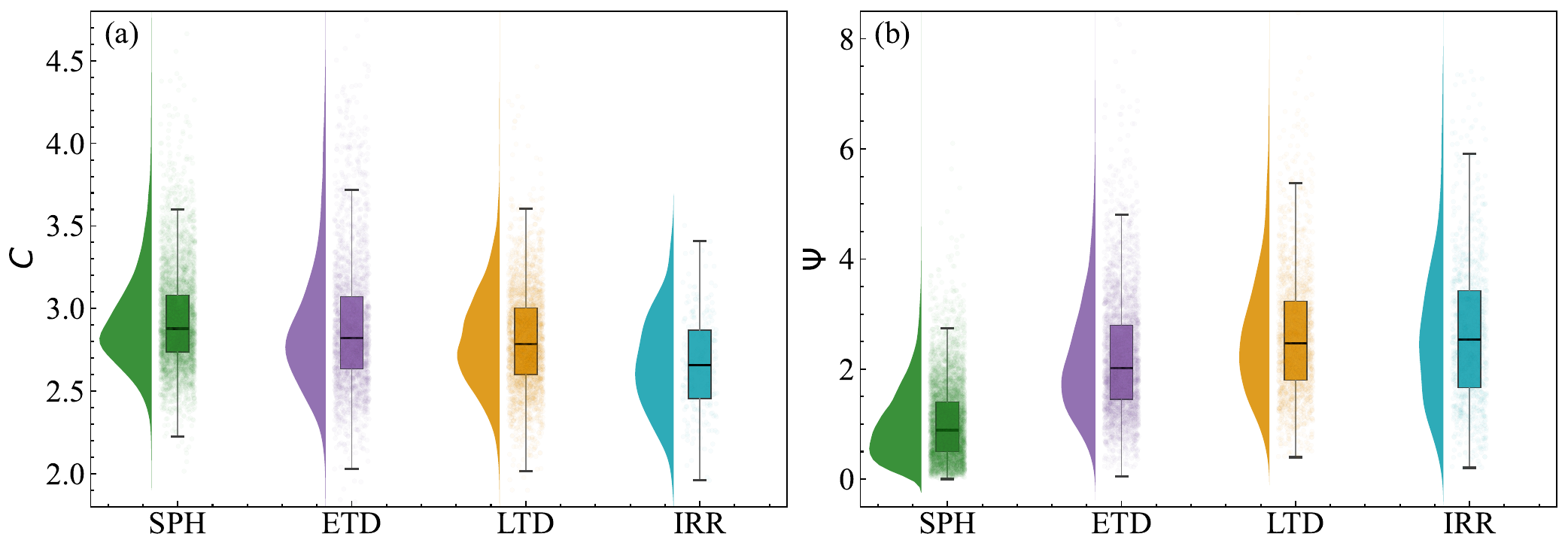}    
\caption{Raincloud plots of C (left) and $\Psi$ (right) for different types of massive galaxies. Plots are analogous to those in Fig.~\ref{fig:8}. From SPHs to IRRs, C decreases while $\Psi$ gradually increases.}    
\label{fig:10}
\end{figure*}

\subsubsection{Parametric measurements}
We used the {\tt\string GALFIT} package \citep{Peng+2002} and {\tt\string GALAPAGOS} software \citep{Barden+2012, Haussle+2022} to measure the morphological parameters of galaxies. We fit the 2D surface-brightness profiles of each galaxy through an optimized single-component S\'{e}rsic model, from which we derived two fundamental morphological parameters: the S\'{e}rsic index ($n$), which characterizes the concentration of the light profile, and the effective radius ($r_e$), the radius at which half of the total light from the galaxy is emitted.

Fig.~\ref{fig:8}a shows the distribution of the S\'{e}rsic index for the four types of galaxies. The median values of the S\'{e}rsic index for SPHs, ETDs, LTDs, and IRRs are 2.16, 1.72, 1.22, and 1.21, respectively. In general, the S\'{e}rsic index of elliptical galaxies is greater than 2, whereas that of disk galaxies is smaller than 2 \citep{fisher+2008,blanton+2009}. Fig.~\ref{fig:8}b shows the distribution of the effective radii of the four galaxy types, with median values of 1.21, 1.79, 2.74, and 2.21 kpc for SPHs, ETDs, LTDs, and IRRs, respectively. Overall, the distributions and trends of the S\'{e}rsic index and effective radii for different types of galaxies are consistent with the expected relationships for galaxy types in the structural parameter space. The larger $r_e$ observed for LTDs compared to IRRs may be attributed to the fact that the selected galaxy sample predominantly consists of low-mass systems.

\subsubsection{Nonparametric measurements}
During our morphological analyses, we used \texttt{statmorph\_csst} \citep{Yao+2023} to measure key nonparametric morphological indicators across the galaxy sample. This comprehensive analysis includes the calculation of the Gini coefficient (G), second-order moment of light ($M_{\rm 20}$), and concentration (C). In addition,  we measured the $\Psi$ and MID (multimode, intensity, deviation) parameters \citep{Rodriguez_Gomez+2018}, allowing a multidimensional characterization of the galaxy morphology.

The G serves as a robust quantitative measure for characterizing the inequality in flux distribution among a galaxy's pixels, providing valuable insights into its structural properties \citep{Lotz+2004, Lotz+2008}. Mathematically, the G is defined as
\begin{equation}    
G=\frac{1}{\overline{f}n_{pix}(n_{pix}-1)}\sum_{i=0}^n(2i-n_{pix}-1)f_i,
\end{equation}
where $n_{pix}$ represents the number of pixels that make up the galaxy, $f_i$ indicates the pixel flux values sorted in ascending order, and $\overline{f}$ denotes the average flux value across all pixels. The parameter $M_{20}$ is a robust morphological indicator that quantifies the spatial concentration of the brightest regions of a galaxy, used primarily to measure the spatial structure of its brightness distribution \citep{Lotz+2004}. It can be defined as the normalized second-order moment of the brightest 20\% of the galaxy's total flux:
\begin {equation}
\begin{aligned}
&M_{\rm tot}=\sum _{i}^{n}M_{i}=\sum _{i}^{n}f_{i}[(x_{i}-x_{c})^{2}+(y_{i}-y_{c})^{2}];\\
&M_{\rm 20} = \log_{10}\frac{\sum_i M_i} {M_{\rm tot}}, \ 
\mathrm{while}\ 
\sum_if_i < 0.2 f_{\rm tot},
\end{aligned}
\end{equation}
where ${M_i}$ represents the second-order moment of pixel $i$, $f_i$ denotes the flux at $i$-th pixel, $(x_i,\ y_i)$ signifies the position of pixel $i$, $(x_c,\ y_c)$ are the coordinates of the galaxy's center, $M_{\rm tot}$ is the total second-order moment of all pixels, and $f_{\rm tot}$ represents the total flux of the galaxy \citep{Lotz+2004, Lotz+2008}.

Figure ~\ref{fig:9}a shows the distribution of G for the four types of galaxies, with median values of SPHs, ETDs, LTDs, and IRRs of 0.52, 0.51, 0.50, and 0.49, respectively. Figure ~\ref{fig:9}b shows the $M_{\rm 20}$ distributions, which increase from SPHs to IRRs, with median values of -1.75, -1.73, -1.64, and -1.44, respectively. These distributions of $G$ and $M_{\rm 20}$ for these four types of galaxies are consistent with previous results \citep{2021MNRAS.503.4446C, Dai+2023, Song+2024}.

The concentration (C) quantifies the central light distribution  relative to the outer regions \citep{Conselice+2000, Conselice+2003}. It is defined as
\begin {equation}
C = 5 \log_{10} \left( \frac{r_{80}}{r_{20}} \right),
\end{equation}
where $r_{20}$ and $r_{80}$ represent the radii of circular apertures containing 20\% and 80\% of the galaxy’s light, respectively \citep{Conselice+2003}. In contrast,  $\Psi$ serves to illustrate how the light distribution of galaxies can be broken down into apparent components \citep{Law+2007}. It is defined as
\begin{equation}
\Psi = 100 \log_{10} \left( \frac{\psi_{\text{compact}}}{\psi_{\text{actual}}} \right),
\end{equation}
where $\psi_{\text{compact}}$ is the sum of the projection potentials in the most compact circular configuration, with pixels arranged from the brightest at the center to decreasing fluxes outward; $\psi_{\text{actual}}$ is the sum of the projection potentials based on the actual distribution of the image, accounting for interactions between all pairs of pixels. This formula quantitatively compares the actual image distribution with an idealized compact model, providing a measure of how much the real distribution deviates from the most compact configuration.

\begin{figure*}[ht!]    
\includegraphics[width=2\columnwidth]{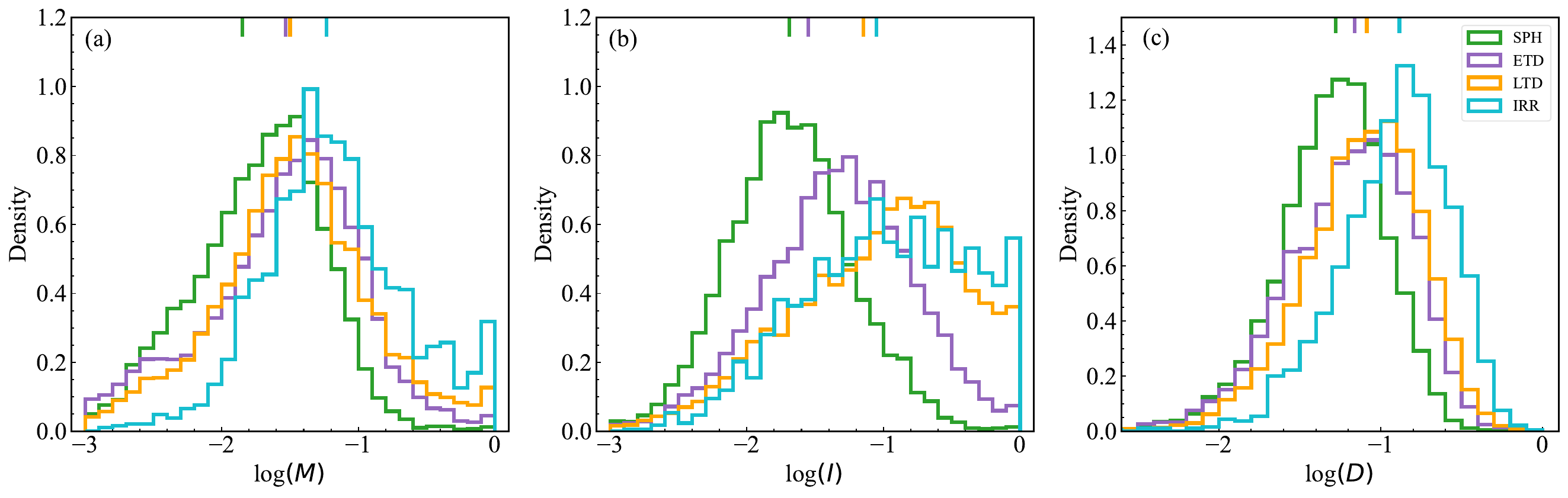}    
\caption{Histogram distributions of multimode (M, panel a), intensity (I, panel b), and deviation (D, panel c) for different types of massive galaxies. Color bars at the top of each panel show the median values of M, I, and D for each galaxy type. Notably, the MID median values gradually increase from SPH to IRR types, reflecting a systematic trend of different galaxy types in the MID parameter space.}    
\label{fig:11}
\end{figure*}

Figure~\ref{fig:10}a shows that more compact SPHs have larger C values than IRRs, displaying a decreasing trend from SPHs to IRRs, with median values of 2.88, 2.82, 2.78, and 2.66, respectively. This is consistent with previous studies showing that more compact galaxies have larger C values \citep{Ferreira+2023,Song+2024, Fang+2024, Tohill+2024}.
Figure~\ref{fig:10}b shows the distribution of the $\Psi$ parameter, displaying an increasing trend from SPHs to IRRs, and median values of $\Psi$ for SPHs, ETDs, LTDs, and IRRs of 0.89, 2.01, 2.47, and 2.53, respectively.
These results are consistent with previous studies, supporting the reliability of our $\Psi$ and $C$ nonparametric measurements.

The MID system is a sophisticated quantitative framework for galaxy morphology analysis, providing enhanced identification capabilities beyond traditional classification schemes \citep{Freeman+2013, Rodriguez_Gomez+2018}. This triparametric system characterizes galaxy structure through three complementary metrics. The metric $M$ evaluates the prominence of the two most dominant clumps within a galaxy by assessing their area ratio. The computation involves identifying these substructures through the following steps. Initially, pixels from the galaxy segmentation map are ordered according to their brightness levels. For a given quantile $q$ (ranging from 0 to 1), all pixels exceeding the brightness threshold defined by the $q$ quantile are grouped into contiguous sets, which are then ordered by area in descending order: $A_{q,1}, A_{q,2}, \ldots, A_{q,n}$. The value of $M$ corresponds to the quantile $q$ that achieves the highest ratio of the second-largest area to the largest:
\begin{equation}
M = \max_q \left( \frac{A_{q,2}}{A_{q,1}} \right).
\end{equation}
Therefore, $M$ identifies the quantile that maximizes the proportion of the second-largest area relative to the largest.

The metric $I$ quantifies the relative brightness between the two most luminous segments within a galaxy \citep{Freeman+2013, Rodriguez_Gomez+2018}. It is formulated as
\begin{equation}
I = \frac{I_2}{I_1},
\end{equation}
where $I_1$ and $I_2$ denote the intensity of the brightest and second brightest segments. 
This approach enables the evaluation of contrast between the foremost bright areas within the galaxy's structure.

The metric $D$ measures the separation between the center of mass $(x_c, y_c)$ of the pixels identified through MID segmentation and the location of the brightest peak $(x_{I_1}, y_{I_1})$, which is determined during the calculation of the intensity statistic $I$. This distance is then normalized by dividing it by the square root of the ratio of the number of pixels within the segment to $\pi$. Specifically, $n_{\text{seg}}$ represents the pixel count within the segmentation area, serving as an approximation of the galaxy's radius \citep{Freeman+2013, Rodriguez_Gomez+2018}:
\begin{equation}
D = \sqrt{\frac{\pi}{n_{\text{seg}}}} \cdot \sqrt{(x_c - x_{I_1})^2 + (y_c - y_{I_1})^2}.
\end{equation}
This formulation effectively captures the relative displacement between the galaxy's centroid and its most luminous point, adjusted for the galaxy's size.

Figure~\ref{fig:11} illustrates that as the luminosity contrast between two regions within galaxies intensifies and the distance from the image's light-weighted center to the brightest peak grows, the values of $M$, $I$, and $D$ also increase. Specifically, SPH galaxies exhibit lower values for $M$, $I$, and $D$. In contrast, ETD, LTD, and IRR galaxies exhibit higher values of these metrics. This clear distinction among different galaxy types in parameter space confirms the reliability of our classification approach.

\section{Summary} \label{sec:5}
This work improves the UML process of the two-step galaxy morphological classification framework by incorporating contrastive learning and PCA to enhance feature extraction and dimensionality reduction. 
Based on this improved framework, 46,176 galaxies at $0<z<4.2$ selected from the COSMOS-Web field are classified into five types of galaxies. This classification system operates through four key steps. First, original galaxy images undergo preprocessing using CAE and APCT. This preprocessing serves to mitigate noise and enhance the rotational invariance of the images. Second, the preprocessed images are encoded using pre-trained ConvNeXt and ViT models. Subsequently, the features are further refined and downscaled via contrastive learning and PCA. Third, during the UML process, 46,176 galaxies from the COSMOS-Web field were clustered using a bagging-based model. Among these, 17,326 galaxies were successfully grouped into 20 clusters, which were then visually classified into five morphological classes. Fourth, the labels of the successfully classified galaxies were utilized to train the SML model (GoogLeNet) for the 28,850 galaxies that were excluded during the UML process. Ultimately, a total of 46,176 galaxies were categorized into 15,191 SPHs, 11,439 ETDs, 11,614 LTDs, 3,121 IRRs, and 4,811 UNCs.

The inclusion of contrastive learning and PCA enhances the model's ability to extract crucial features and enables the acquisition of a galaxy sample with high-confidence labels during the UML phase. As a result, training the SML model with this labeled galaxy image dataset significantly improves the prediction accuracy for the galaxies discarded during UML. This enhancement allows the system to manage complex astronomical datasets more effectively, leading to improved classification accuracy and providing a more reliable method for large-scale galaxy morphology classification.

\section*{Data availability statement}
The catalog of all galaxy samples in this paper is only available in electronic form at the CDS via anonymous ftp to cdsarc.u-strasbg.fr (130.79.128.5) or via \url{ http://cdsweb.u-strasbg.fr/cgi-bin/qcat?J/A+A/.}

\begin{acknowledgements}
S.L. acknowledges the support from the Key Laboratory of Modern Astronomy and Astrophysics (Nanjing University) by the Ministry of Education. Z.S.L. acknowledges the support from Hong Kong Innovation and Technology Fund through the Research Talent Hub program (PiH/022/22GS). The numerical calculations in this paper have been done on the computing facilities in the High Performance Computing Platform of Anqing Normal University.
\end{acknowledgements}

\bibliographystyle{aa}
\bibliography{ref}
\end{document}